# Atomic-Scale Observation of Symmetry Breaking in Altermagnetic MnTe

Guodong Ren[1,*], Jonathan M. DeStefano[2], Xiao-Wei Zhang[1], Arashdeep S. Thind[3], Rajiv Giridharagopal[4], José Ángel Castellanos-Reyes[5], Paul M. Zeiger[1,5], Noah Kamm[3], Sijie Xu[6,7], Zhaoyu Liu[6,7], Yaofeng Xie[2], Filip Krizek[8], Jan Michalička[8,9], Richard Campion[10], Pengcheng Dai[6,7], Peter Wadley[10,11], David S. Ginger[4], Tomas Jungwirth[8,10,11], Robert F. Klie[3], Ján Rusz[5], Di Xiao[1,2], Jiun-Haw Chu[2], Juan Carlos Idrobo[1,12,*]

[1] Department of Materials Science and Engineering, University of Washington, Seattle, WA 98195, USA
[2]Department of Physics, University of Washington, Seattle, WA 98195, USA
[3]Department of Physics, University of Illinois Chicago, Chicago, IL 60607, USA
[4]Department of Chemistry, University of Washington, Seattle, WA 98195, USA
[5]Department of Physics and Astronomy, Uppsala University, Uppsala, Box 516, 75120, Sweden
[6]Department of Physics & Astronomy, Rice University, Houston, TX 77005, USA
[7]Rice Laboratory for Emergent Magnetic Materials and Smalley-Curl Institute, Rice University, Houston, TX 77005, USA
[8]Institute of Physics, Czech Academy of Sciences, Cukrovarnicka 10, 162 00 Prague 6, Czech Republic
[9]Central European Institute of Technology, Brno University of Technology, Purkynova 123, 612 00 Brno, Czech Republic
[10]School of Physics and Astronomy, University of Nottingham, Nottingham NG7 2RD, United Kingdom
[11]Laboratory for Nanoelectronics and Spintronics, Tohoku University, 2-1-1 Katahira, Aoba-ku, Sendai 980-8577, Japan
[12]Physical and Computational Sciences Directorate, Pacific Northwest National Laboratory, Richland, WA 99354, USA

*Corresponding authors: gdren@uw.edu, jidrobo@uw.edu

## Abstract

**The recent discovery of altermagnetism has sparked growing interest in compensated magnetic systems as promising platforms for highly scalable spintronics. Altermagnetism is a distinct magnetic order where opposite spin sublattices are connected by rotation, yielding zero net magnetization but momentum-dependent spin splitting. To date, experimental verification of altermagnetic order has been achieved predominantly through bulk-sensitive techniques, including spin-dependent electronic spectra and transport responses. However, direct atomic-scale evidence that explicitly correlates crystal symmetry, local structural distortions, and magnetic ordering has remained unexplored. Here, we report the direct atomic-scale observation of coexisting polar distortions and altermagnetic order in MnTe, combining atomic-resolution scanning transmission electron microscopy (STEM) imaging with electron magnetic chiral dichroism (EMCD) measurements. We reveal that MnTe is not an ideal uniform $P6_3/mmc$ *g*-wave altermagnet at the atomic scale. Instead, it hosts ubiquitous inversion-symmetry-breaking distortions that lower the spin-space-group (SSG) symmetry, admits *d*-wave altermagnetic components, and in lower-symmetry regimes, even allow *s*-wave spin splitting (net magnetization). The coexistence of ferroelectric signatures and altermagnetic order establishes local lattice symmetry in MnTe as a control knob for altermagnetic spin splitting, spin current generation, and multiferroic memory applications.**

Understanding the symmetry properties and their associated order parameters of crystalline solids is central to modern condensed matter physics. The intimate interplay between spin, orbital and lattice degrees of freedom has led to a broad range of exotic phases and emergent phenomena. Enabled by its unique magnetic crystal symmetry, altermagnetism has emerged as a newly recognized class of collinear magnetic order characterized by complete spin compensation, yet hosting strong momentum-dependent spin splitting and time-reversal symmetry breaking in its electronic bands, even in the absence of spin–orbit coupling and net magnetization[1-5]. In collinear antiferromagnets, where the spin sublattices are connected by combining time-reversal ($T$) and a half-translation ($t_{1/2}$) or inversion ($P$), spin degeneracy in momentum space is generally protected by $T\cdot t_{1/2}$ symmetry in the absence of spin–orbit coupling (SOC), whereas $P\cdot T$ symmetry preserves degeneracy even in the presence of SOC. Altermagnets, in contrast, connect opposite spin sublattices with a rotation transformation (proper or improper), which simultaneously break the $P\cdot T$ and $T\cdot t_{1/2}$ symmetries, leading to non-relativistic momentum-dependent spin splitting despite the absence of net magnetization. Beyond its fundamental significance for unconventional band splitting, altermagnetic order can host a variety of $T$-symmetry breaking responses traditionally ascribed to the macroscopic magnetization in ferromagnets, including the anomalous Hall effect[6-8], spin current generation[9-11], or spin transfer torque[12,13], while retaining the advantages of vanishing stray field and fast switching dynamics typically associated with antiferromagnets[14].

Altermagnetism has been predicted in a broad range of 2-dimensional and 3-dimensional material systems that were previously classified as antiferromagnets[2,5,15,16]. A prototypical example is hexagonal manganese telluride (α-MnTe), with crystallographic space group $P6_3/mmc$, which has been extensively explored both theoretically and experimentally. Evidence for the altermagnetic character in α-MnTe includes the observation of the spontaneous anomalous Hall effect (AHE)[17-19], which indicates $T$-symmetry breaking in the electronic band structure. In addition, pronounced anisotropic band splitting in momentum space, supported by first-principles calculations and direct angle-resolved photoemission spectroscopy (ARPES) measurements, further substantiates its altermagnetic nature[20-22].

Recent studies also demonstrate the existence of a nonzero magnetization in both epitaxial thin films and freestanding bulk MnTe, attributed to a slight magnetic canting that departs from the established collinear compensated magnetic order[23-25]. The resulting weak ferromagnetic moment was measured to be on the order of $\sim 10^{-5}$ to $10^{-3}$ $\mu_B$/Mn, oriented parallel to the $c$-axis even in the absence of an external magnetic field[23,24]. In α-MnTe, the non-relativistic spin-space symmetry enforces the compensation between magnetic sublattices, prohibiting net magnetization, whereas the relativistic magnetic space-group (MSG) symmetry allows magnetic canting and thus a net moment. The weak ferromagnetism measured in MnTe has therefore been explained by microscopic mechanisms, involving higher-order SOC effects or Dzyaloshinskii–Moriya (DM) type interactions[23,26]. These analyses generally assume a centrosymmetric $P6_3/mmc$ crystal structure, without considering any symmetry-allowed structural distortions. However, recent Raman and IR spectra measurements of α-MnTe reveal multiple phonon modes beyond those expected from $P6_3/mmc$ symmetry[27,28]. Moreover, the second-harmonic generation (SHG) measurements also indicate inversion-symmetry-breaking in α-MnTe[28].

Here, we performed STEM imaging combined with electron energy-loss spectroscopy (EELS) measurements to resolve the local structure and in-plane magnetic order of α-MnTe at the atomic-scale. In our quantitative STEM image analysis, we did not observe the nominal $P6_3/mmc$ crystal structure in any of the bulk crystals or thin films studied, but instead we identified ubiquitous inversion-symmetry-breaking arising from non-centrosymmetric displacements of the Mn and Te atoms. Group-theoretical analysis further reveals that these atomic displacements correspond to the $\Gamma$-point distortion modes inherited from the parent $P6_3/mmc$ space group. The dominant Mn-site distortions condense into lower-symmetry structures of $Cmc2_1$ and $Amm2$, both of which are compatible with non-relativistic $d$-wave altermagnetic order described by SSG $C^{1}m^{\bar{1}}c^{\bar{1}}2_1^{\infty_{010}m}1$ and $A^{\bar{1}}m^{1}m^{\bar{1}}2^{\infty_{001}m}1$, respectively. Both motifs carry a common $E_{2g}$ vestigial orthorhombic order parameter, providing a unifying structural-symmetry-breaking channel consistent with the recently reported detwinning of altermagnetic domains in MnTe under in-plane uniaxial strain[18]. Moreover, the combined Mn- and Te-sublattice distortions yield a lower-symmetry $Cm$ structure that removes the exact altermagnetic symmetries, producing a mixed $d$-wave plus $s$-wave state. Quantitative EMCD analysis based on the Mn $L_{2,3}$ edge reveals a periodic magnetic chiral dichroic signal across successive Mn layers in $Cmc2_1$ and $Amm2$ lattices, consistent with dominant in-plane alternating magnetic order in the distorted structures as demonstrated in our density-functional theory (DFT) calculations. Furthermore, piezoresponse force microscopy (PFM) measurements capture the ferroelectric signatures of the inversion-symmetry-breaking polar states, opening the possibility of efficient electric-field control of the altermagnetic order.

Figure 1**a** shows a schematic crystal of α-MnTe with $P6_3/mmc$ symmetry. This structure is composed of face-shared $MnTe_6$ octahedra running along the $c$-axis. Below the Néel temperature, the magnetic moments on Mn align antiparallel between two spin sublattices ($Mn_A$ and $Mn_B$) within the $a$-$b$ plane. The $Mn_A$ and $Mn_B$ spin sublattices are not connected by translation or inversion operations but by a twofold rotation in spin space combined with a sixfold screw rotation in real space ($[C_2||C_6t_{1/2}]$), giving rise to $g$-wave altermagnetism.

In this work, bulk single-crystalline α-MnTe was synthesized using the molten flux method (see more details in *Methods*). Temperature-dependent resistivity measurements (see Figure S2 in *Supplementary Materials*) determine the Néel temperature $T_N \approx 305$ K, in good agreement with values previously reported in the literature[29,30]. The MnTe thin film was grown by molecular beam epitaxy (MBE) on a single-crystalline InP(111) substrate, of which the altermagnetic order features have been confirmed by ARPES[20] and X-ray magnetic circular dichroism (XMCD) measurements[24].

Although the structural symmetry of MnTe crystals has generally been determined to be the centrosymmetric space group $P6_3/mmc$, recent Raman spectroscopy and second-harmonic generation (SHG) measurements reveal anomalous phonon and optical activities that violate the $P6_3/mmc$ symmetry[27,28]. To resolve structural ambiguities, we performed atomic-resolution STEM imaging to probe the atomic-scale crystal structure of MnTe and to identify any local structural distortions or defects associated with symmetry breaking. Figure 1**b-c** present the representative high-angle annular dark field (HAADF) STEM images of the α-MnTe lattice viewed along the [100] and [001] zone axes, respectively. The heavier Te atomic columns appear brighter than the lighter Mn columns due to *Z*-contrast characteristic of HAADF imaging[31]. In order to determine the elemental distribution, we performed atomically resolved EELS mapping of the core-level excitations based on the characteristic edges of Te $M_{4,5}$ and Mn $L_{2,3}$. The results are shown in the right panel of Figure 1**b-c** and verify that Te and Mn atoms are occupying the expected crystallographic sites.

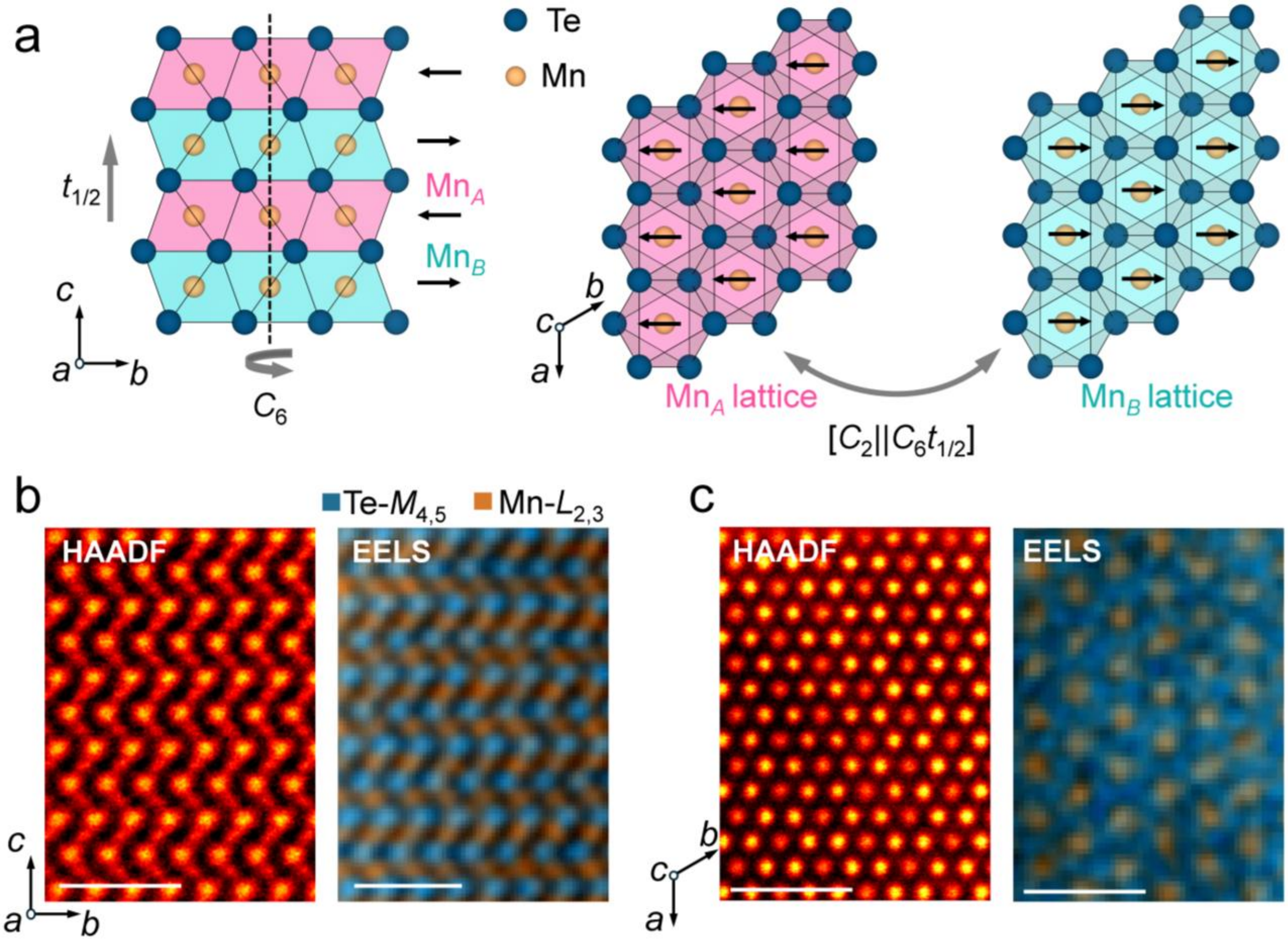

**Figure. 1 Crystallographic and magnetic structure of α-MnTe**. **(a).** The schematic lattice of α-MnTe having the centrosymmetric space group $P6_3/mmc$, in which face-sharing $MnTe_6$ octahedra are stacked along the *c*-axis and connected by the screw operation, $C_6t_{1/2}$. The magnetic moments are antiparallel in alternating Mn layers ($Mn_A$ and $Mn_B$ sublattices), leading to antiferromagnetic order that lowers the magnetic symmetry to the magnetic space group *Cm'c'm*. **(b).** Left panel: A representative STEM-HAADF image showing the local atomic structure viewed along the *a*-axis. Right panel: STEM-EELS mapping showing the spatial distribution of the integrated Te-$M_{4,5}$ and Mn-$L_{2,3}$ signals. **(c).** The representative STEM-HAADF image (left panel) and STEM-EELS mapping (right panel) showing the local atomic structure and elemental distribution viewed along the *c*-axis. Scale bars are 1 nm in **(b)** and **(c)**.

To directly visualize the local structural order associated with atomic displacements, we quantified the position offsets of Mn and Te columns recorded in STEM images. Using the large field-of-view HAADF-STEM image shown in Figure 2**a**, the positions of Te and Mn atomic columns were extracted with subpixel precision by fitting two-dimensional Gaussians to each atomic column (see Figure S3 in *Supplementary Materials*). We observed off-center displacements of the Mn atomic columns relative to the centroid of their coordinating Te columns. These displacements within $MnTe_6$ octahedra were quantified across the entire region shown in Figure 2**a** and are overlaid as displacement vectors on a vector-color map shown in Figure 2**b**. The displacement vector color map shows that the orientations of Mn distortions are locally aligned but segmented over longer length scales, giving rise to domain-like structures.

Two dominant distortion patterns associated with the Mn displacements are identified and visualized in magnified views in Figure 2**e** and Figure 2**g**, respectively. The local structural motif of α-MnTe shown in Figure 2**e** reveals alternating Mn layers undergoing opposite displacements along the horizontal direction, while collectively shifting upward along the vertical direction, corresponding locally to a polar displacement pattern along the *c*-axis. A statistical analysis of the corresponding Mn-layer displacements is presented in Figure 2**f**: the horizontal components ($u_H$) alternate between positive and negative values, indicating rightward and leftward shifts, whereas the vertical components ($u_V$) are exclusively positive, consistent with a collective upward displacement. The polar displacements of Mn atoms along the *c*-axis can occur either upward or downward, resulting in the formation of opposing polar domains (see Figure S4 in *Supplementary Materials*). Figure 2**g** displays another structural motif, which features alternating Mn layers undergoing opposite displacements along the vertical direction while collectively shifting rightward horizontally, corresponding locally to an in-plane polar displacement pattern. The corresponding statistical analysis in Figure 2**h** quantifies the Mn displacements associated with this structural motif.

In addition to the Mn off-center displacements within $MnTe_6$ octahedra, we also quantified the Te sublattice displacements. The nearest-neighbor Te–Te distances were measured along the vertical ($d_V$) and horizontal ($d_H$) directions, respectively. As shown in Figure 2**c**, the vertical Te–Te spacing exhibits clear modulation, corresponding to the breathing-type distortions along the *c*-axis. In contrast to the pronounced domain-like Mn displacements observed in Figure 2**b**, the Te sublattice displacements display a more diffuse and spatially gradual variation across the field of view. This behavior suggests that the Te and Mn sublattice distortions are not strongly correlated. The Te–Te spacing along the horizontal direction on the other hand is rather uniform across the field of view (see Figure S5 in *Supplementary Materials*). Correspondingly, the histogram in Figure 2**d** reveals a bimodal distribution for the vertical Te–Te distances, whereas the horizontal Te–Te distances show a unimodal distribution. The more diffuse modulation of the Te–Te spacing compared with the domain-like Mn displacement field, suggests that the inversion-breaking states in MnTe may not be described by a single local polar distortion pattern involving both sublattices equally.

To verify the ubiquity of inversion-breaking distortions in α-MnTe, we also performed atomic resolution STEM imaging on a thin film sample epitaxially grown on InP(111) substrate (see Figure S7 and S8 in *Supplementary Materials*). Consistent with the bulk sample, non-centrosymmetric distortions involving both Mn and Te displacements were also identified in the thin films.

In order to determine the symmetry of the associated atomic distortions observed in the experimental images, we applied group-theoretical analysis to identify the symmetry-allowed subgroup distortions of the $P6_3/mmc$ structure. Figure 2**i-n** illustrate symmetry-relevant zone-center distortion modes associated with the observed Mn and Te displacements in α-MnTe structure (volume-changing modes involving expansion or shearing distortion are neglected), with colored arrows indicating the displacement directions. We notice that a combination of the out-of-plane polar $\Gamma_2^-$ and in-plane antipolar $\Gamma_5^-$ modes lead to the formation of the structural motif shown in Figure 2**e**. While the motif in Figure 2**g** results from combining the out-of-plane antipolar $\Gamma_4^-$ with the in-plane polar $\Gamma_6^-$ modes. Meanwhile, the $\Gamma_3^+$ mode corresponds to the breathing-type displacements of the Te sublattice along the *c*-axis, as observed in Figure 2**c**.

Symmetry analysis indicates that the combined $\Gamma_2^-$ and $\Gamma_5^-$ modes lower the symmetry to $Cmc2_1$, whereas the combination of $\Gamma_4^-$ and $\Gamma_6^-$ mode results in $Amm2$ symmetry (see Figure S9 in *Supplementary Materials*). Considering the possible coexistence of Mn and Te displacements, the symmetry can be further reduced to $Cm$. Figure S11 schematically shows opposite spin sublattices within $Cmc2_1$ and $Amm2$ structural motifs connected by rotational operations $[C_2||C_2t_{1/2}]$ and $[C_2||C_2]$, respectively. Within the symmetry framework of SSG[32-36], the non-centrosymmetric $Cmc2_1$ and $Amm2$ lattices can respectively host non-relativistic SSG symmetries $C^1m^{\bar{1}}c^{\bar{1}}2_1^{\infty_{010}m}1$ and $A^{\bar{1}}m^1m^{\bar{1}}2^{\infty_{001}m}1$ (here we follow the Chen-Liu SSG notation[32,36]), both of which are compatible with *d*-wave altermagnetism.

Despite arising from distinct *u*-mode combinations, the $Cmc2_1$ and $Amm2$ motifs share a common secondary feature. The bilinear products of the condensing modes both transform as the in-plane orthorhombic representation $E_{2g}$ of the parent $P6_3/mmc$: $A_{2u}\,(\Gamma_2^-) \otimes E_{2u}\,(\Gamma_5^-) = E_{2g}\,(Cmc2_1)$ and $B_{2u}\,(\Gamma_4^-) \otimes E_{1u}\,(\Gamma_6^-) = E_{2g}\,(Amm2)$. Each motif therefore carries a common vestigial $E_{2g}$ order parameter — an in-plane orthorhombic distortion that lowers the 6-fold rotational symmetry of the parent lattice to 2-fold. The $E_{2g}$ component provides a unifying structural order parameter across the two motifs, predicts a measurable in-plane orthorhombic distortion, and offers a natural explanation for the recent observation that in-plane uniaxial strain detwins the three 120°-related altermagnetic domains in α-MnTe[18].

In contrast to the $Cmc2_1$ and $Amm2$ lattices, the $Cm$ lattice hosts oppositely oriented spin sublattices that cannot be interconverted by translation (or inversion) nor by rotation operations. The $Cm$ structure breaks the symmetries of an exact altermagnetic order, yet its SSG $C^1m^{\infty_{100}m}1$ still permits non-relativistic spin splitting. Moreover, the spin-only group of this SSG is the polar group $\infty m$ (with continuous rotations about [100] and a mirror plane containing that axis), which permits a non-zero net magnetization ($s$-wave) along the collinear axis [100] even without SOC. However, the experimentally reported weak moment in α-MnTe is out-of-plane (along the $P6_3/mmc$ $c$-axis)[23,24], thus the non-relativistic $s$-wave state in $Cm$ cannot directly account for the observed out-of-plane weak ferromagnetism—that component still requires SOC. The $Cm$ lattice could instead host a separate in-plane non-relativistic magnetization that has not yet been experimentally sought. As demonstated in Figure S13, with SOC involved, SSGs lower symmetry to the relativistic MSGs, which allow the out-of-plane canted moments in $Cmc2_1$, $Amm2$ and $Cm$ lattices —all of which can contribute to the out-of-plane weak ferromagnetism.

The observed inversion-symmetry-breaking distortions could naturally explain the intrinsic non-centrosymmetric distortions recently revealed by Raman spectroscopy and SHG measurements[27,28]. They also enable ferroelectric order alongside altermagnetism in MnTe. PFM measurements on a bulk α-MnTe crystal (see **Section 5** in *Supplementary Materials*) reveal well-defined domains with clear phase contrast (Figure S10**d**), showing negligible correlation with the topographic features in Figure S10**b**. Bright phase contrast regions exhibit hysteresis as well as amplitude "butterfly" loops (Figure S10**e**), characteristic of ferroelectric domains, and likely correspond to $Cmc2_1$ or $Cm$ lattices with dominant out-of-plane polarization. In contrast, low-contrast regions show almost no hysteresis (Figure S10**f**), suggesting that the $Amm2$ lattice with in-plane polar displacements is possibly dominant.

To gain insights into the effect of structural distortions on spin splitting, we carried out first-principles DFT calculations on the experimentally observed structures with inversion-symmetry-breaking. At $k_z = 0$ (use the $P6_3/mmc$ $k$-space coordinates for all lattices), with SOC turned off, the spin degeneracy along the $-K-\Gamma-K$ nodal line is protected by the SSG symmetry operations for $P6_3/mmc$, $Amm2$ and $Cmc2_1$ lattices, as shown in Figure S11. While at $k_z \neq 0$, non-relativistic spin splitting emerges along -$\bar{M}-\bar{\Gamma}-\bar{M}$ (Néel-vector easy axis), yet degeneracy is retained at $\bar{\Gamma}$ , a hallmark of altermagnetism. Owing to the broken inversion-symmetry, the *d*-wave altermagnetic orders in $Amm2$ and $Cmc2_1$ lattices have two spin-degenerate nodal planes, in contrast to four nodal planes of the *g*-wave order in $P6_3/mmc$ lattice. When altermagnetic symmetries are further broken in *Cm* lattice, besides having the non-relativistic spin splitting along both -$\bar{M}-\bar{\Gamma}-\bar{M}$ and -$K-\Gamma-K$, spin degeneracy is also lifted at $\Gamma$ and $\bar{\Gamma}$, indicating a mixed *d*-wave and *s*-wave state.

With SOC taken into account (see Figure S12 in *Supplementary Materials*), the relativistic spin splitting occurs along both -$\bar{M}-\bar{\Gamma}-\bar{M}$ and -$K-\Gamma-K$ in all lattices. Moreover, due to the interplay of inversion-symmetry-breaking and SOC, the spin-resolved band dispersions become asymmetric between positive and negative momenta in the noncentrosymmetric lattices.

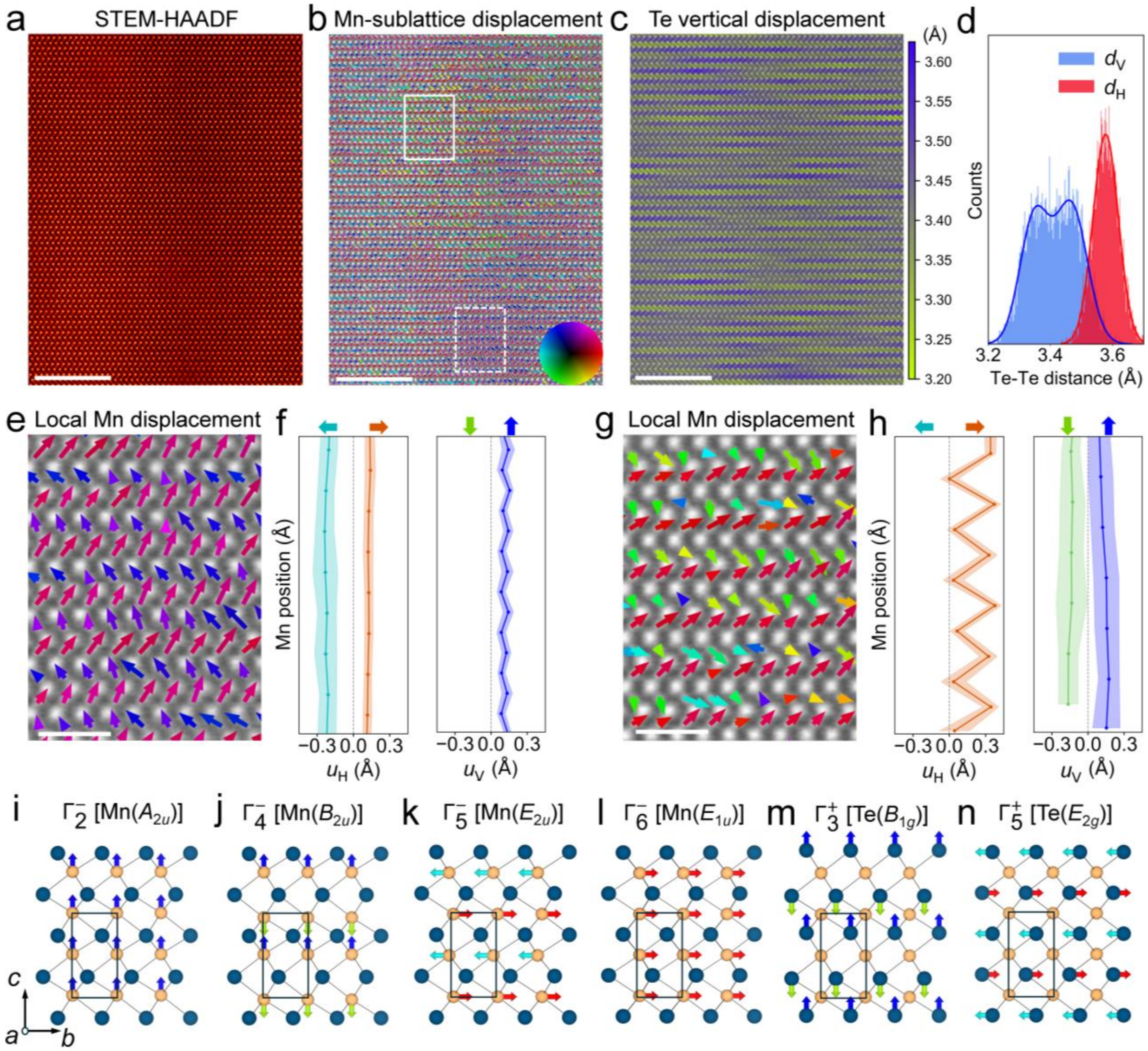

**Figure. 2 Atomic-scale structural determination and local symmetry breaking in MnTe**. **(a).** A representative STEM-HAADF image showing the atomic structure of MnTe along the *a*-axis over a large field of view. **(b).** Displacement vector map of the Mn sublattice. The vectors are calculated from the off-centering of Mn atomic columns relative to the centroid of the surrounding Te sublattice and are overlaid on the original HAADF image shown in **(a)**. The direction and magnitude of Mn displacements are represented by arrows. The direction is also indicated by the color wheel for easier visualization. **(c).** Nearest Te-Te distance mapping along the vertical direction ($d_V$) in the original HAADF image shown in **(a)**, which corresponds to displacements of the Te-sublattice along the *c*-axis. **(d).** Histogram plot showing the distribution of Te–Te distances along the vertical ($d_V$) and horizontal ($d_H$) directions in the original HAADF image shown in **(a)**. **(e, g).** Local structural motifs of MnTe extracted from the image shown in **(b)** as indicated by the dashed and solid white frames, respectively. Colored arrows overlaid on the atomic columns indicate the displacement directions and magnitudes of the Mn sublattice. **(f, h).** Statistical analysis of Mn-layer displacements, extracted from the local structural motifs shown in **(e)** and **(g)**, respectively. Displacements of each Mn layer are resolved along the horizontal ($u_H$) and vertical ($u_V$) directions, respectively. Shaded regions in the plots indicate the corresponding standard deviations. **(i-n).** Atomic models showing the possible Γ-point (zone center) distortion modes allowed by the $P6_3/mmc$ space group symmetry. For simplicity, volume-changing modes involving expansion or shearing distortion are neglected. The unit cell of each distortion mode is enclosed by the black solid

lines. Figures (**i-n**) show symmetry-allowed distortion modes used to interpret the locally observed motifs in **e** and **g**. Scale bars are 5 nm in **(a-c**) and 1 nm in **(e)** and **(g)**.

To experimentally verify the preserved altermagnetic order in the low-symmetry polar MnTe lattices, it is necessary to probe the magnetic moment related behavior around individual Mn columns. Electron magnetic chiral dichroism (EMCD)[37], an analog to X-ray magnetic circular dichroism (XMCD), probes electronic transitions associated with spin and orbital angular momenta of electrons and has emerged as an effective tool for quantitative magnetic property measurements with high spatial resolution in STEM. Conventional EMCD measurements require tilting the crystal to a two- or three-beam diffraction geometry, which generates dichroic signals at conjugate scattering angles in the diffraction plane, thereby limiting spatial resolution to the scale of atomic planes[38,39]. In this work, we acquired atomic-scale EMCD signals using on-axis EELS acquisition[40], with an aberration-corrected electron probe as recently demonstrated by Song et al.[41]

For a magnetic Mn atom, where the local time-reversal symmetry is broken, the Mn $L_{2,3}$ edge, corresponding to the electronic transitions from 2*p* to unoccupied 3*d* orbitals, contain both nonmagnetic and magnetic components. As indicated by the EMCD simulations (see Figure S14 in *Supplementary Materials*), when the magnetic moment is oriented perpendicular to the electron beam, the magnetic component of the EELS signals redistributes in real space allowing the detection of magnetic order with atomic resolution[42,43]. Given that the altermagnetic order in α-MnTe consists of predominantly in-plane magnetic moments that are antiparallel between two spin sublattices, we average the Mn $L_{2,3}$ edge signals acquired from $Mn_A$ and $Mn_B$ sublattices to cancel the contribution from in-plane magnetic components, thereby approximating the nonmagnetic contribution under the assumed symmetry of the magnetic configuration (see Figure S16-18 in *Supplementary Materials*). In the present analysis, the experimentally extracted quantity is a layer-resolved dichroic deviation from the average Mn $L_{2,3}$ spectrum of $Mn_A$ and $Mn_B$ sublattices, $\langle \sum (Mn_A+Mn_B) \rangle$, and the interpretation in terms of alternating sublattice magnetism relies on the assumption of a predominant in-plane Mn moment on two opposite spin sublattices.

Figure 3 shows the atomically resolved EMCD signals mapped across two distinct structural motifs in MnTe. The atomic model of *Cmc*2₁-MnTe in Figure 3**a** illustrates the atomic displacements arising from combining $\Gamma_2^-$ and $\Gamma_5^-$ modes. The Mn displacements vectors overlaid on the STEM-HAADF image in Figure 3**c** reflect the corresponding distortion modes of the $Cmc2_1$-MnTe lattice. The STEM-EELS spectra were recorded simultaneously with the STEM-HAADF image, and template-matching was applied to enhance the signal-to-noise ratio (SNR)[44]. Figure 3**e** displays alternating EMCD signals at the $L_3$ edge, extracted from regions offset by 2 pixels above the center of the Mn columns, across the $Mn_A$ and $Mn_B$ sublattices shown in Figure 3**c**, while the dichroic signals at the $L_2$ edge are negligibly small (see Figure S17 in *Supplementary Materials*). According to EMCD sum rules[45,46], the pronounced dichroic response at $L_3$ relative to $L_2$ indicates an enhanced orbital-to-spin magnetic ratio in the distorted $Cmc2_1$-MnTe lattice, which is in agreement with recent theoretical studies predicting a substantial orbital magnetic moment in MnTe[47].

To visualize the spatial variation of EMCD signature associated with altermagnetic order, we mapped the integrated EMCD signals at the $L_3$ edge for each Mn layer, as shown in Figure 3**g**, where opposite signals (highlighted in red and blue) are observed across the $Mn_A$ and $Mn_B$ sublattices. We performed similar measurements and analyses on the *Amm*2-MnTe lattice formed by a combination of the $\Gamma_4^-$ and $\Gamma_6^-$ modes, as shown in Figure 3**b**. The STEM-HAADF image in Figure 3**d** reveals the local structural motif with its characteristic distortion modes present in the *Amm*2-MnTe lattice. As evidenced in Figures 3(**f, h**), EMCD spectra from this lattice also exhibit alternating dichroic signals at the $L_3$ edge, with negligible dichroism at the $L_2$ edge. Therefore, the enhanced orbital-to-spin magnetic ratio is also revealed in the *Amm*2-MnTe lattice. These results support the persistence of the altermagnetic order pattern in both distorted MnTe motifs, with the EMCD signals directly correlating with the dominant in-plane magnetic order on two opposite sublattices.

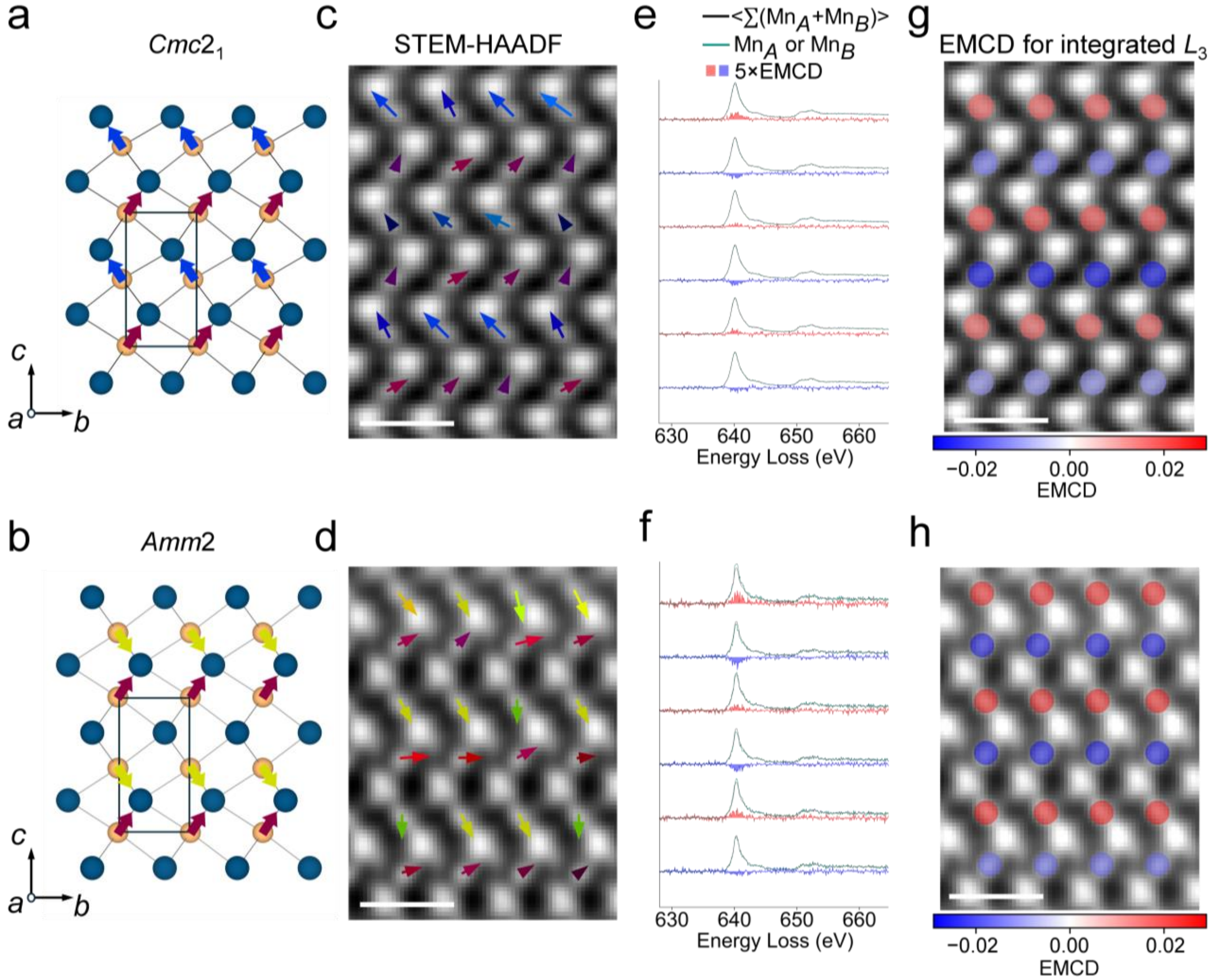

**Figure. 3 Atomic-scale EMCD signatures of magnetic order in local MnTe lattices**. **(a, b).** Atomic models of distorted MnTe phases. The *Cmc*2₁-MnTe model in **(a)** illustrates the atomic displacements arising from combining $\Gamma_2^-$ and $\Gamma_5^-$ modes, and *Amm*2-MnTe in **(b)** displays the combined atomic displacements arising from $\Gamma_4^-$ and $\Gamma_6^-$ modes. **(c, d).** Displacement vector maps of the Mn sublattice corresponding to the local structural motif of $Cmc2_1$-MnTe shown in **(a)** and *Amm*2-MnTe in **(b)**, respectively. **(e, f).** Layer-resolved EELS spectra focusing on Mn $L_{2,3}$ edges, acquired across the local $Cmc2_1$-MnTe structure shown in **(c)** and *Amm*2-MnTe in **(d)**, respectively. The EMCD signals were obtained by taking the difference between the layer-resolved EELS spectra (in green) and the average EELS spectrum over all Mn layers (in black). All spectra shown are background-subtracted and post-edge normalized. **(g, h).** The color and intensity of the circles schematically represent the strength of the EMCD signal at the Mn $L_3$ edge for each atomic row in (**e**, **f**), respectively. Scale bars are 0.5 nm in **(c, d, g, h)**.

MnTe is a polymorphic compound whose phase composition is sensitive to the Te/Mn ratio, growth temperature, and elemental doping[48-50]; therefore, defects or impurity phases can be present in as-grown crystals[51,52]. To evaluate their impact on the altermagnetic order, we characterize the most commonly observed crystal defects in our bulk samples, a half-unit-cell impurity of the monoclinic $P2_1/m$ (MC) phase (see details in **Section 9** of *Supplementary Materials*). Quantitative analysis (Figure S21) reveals three dominant Mn displacement patterns adjacent to the half-unit-cell MC impurity. Moreover, non-centrosymmetric distortions in α-MnTe are found to be sensitive to the local strain field generated by the MC impurity, suggesting that inversion-breaking structural motifs can be tuned by strain-engineering. Furthermore, our EMCD measurements (Figure S24) indicate that altermagnetic order remains preserved up to a few atomic layers away from the MC impurity and is disrupted only in the immediately adjacent layers, underscoring its robustness against local structural perturbations.

We also carried out in-situ heating STEM experiments to determine whether spin-lattice coupling drives the observed structural distortions in MnTe. A microelectromechanical system (MEMS) with in-situ heating E-chip was employed to acquire atomic-resolution STEM images after heating the sample to 100 °C, which is significantly above the Néel temperature (see details in Figure S25-27 of *Supplementary Materials*). The persistence of these distortions and the associated $E_{2g}$ orthorhombicity above $T_N$ indicates that the structural symmetry breaking precedes magnetic ordering, with the in-plane orthorhombic component providing a pre-existing environment that constrains the orientation of the magnetic order parameter once it develops below $T_N$.

In summary, by combining STEM imaging with EMCD measurements, we directly resolved inversion-symmetry-breaking and magnetic ordering in MnTe at the atomic scale. Quantitative STEM data analysis demonstrates that the ideal $P6_3/mmc$ crystal structure is absent from all samples studied. Instead, ubiquitous non-centrosymmetric displacements of Mn and Te atoms produce three distorted lattice motifs with space groups $Amm2$, $Cmc2_1$, and $Cm$, thereby enabling emergent ferroelectric order. Group-theoretical analysis shows that the dominant Mn-site distortions condense into the lower-symmetry $Amm2$ and $Cmc2_1$ structures, both compatible with non-relativistic *d*-wave altermagnetic order, whereas the combined Mn- and Te-site distortions yield a $Cm$ structure that lifts the exact altermagnetic symmetries and hosts a mixed *d*-wave plus *s*-wave state. Moreover, both the $Cmc2_1$ and $Amm2$ motifs carry a common $E_{2g}$ vestigial orthorhombic order parameter, providing a unifying structural-symmetry-breaking channel that couples directly to in-plane uniaxial stress and to altermagnetic domain selection. These observed inversion-symmetry-breaking distortions in α-MnTe naturally reconcile the intrinsic non-centrosymmetric features recently revealed by Raman spectroscopy and SHG measurements[27,28]. Our DFT calculations and EMCD measurements further confirm that these locally broken inversion-symmetry structural motifs preserve the altermagnetic order. By elucidating the interplay between lattice symmetry and magnetic order, this work shows the coexistence of ferroelectricity and altermagnetism in α-MnTe crystal. The two observed coexisting ferroic orders create opportunities for nonvolatile memory device applications, in which electronic properties can be dually controlled by electric fields and spin-dependent phenomena[5,53], however their coupled effect requires further investigations.

## Methods

### Sample growth

The MnTe bulk samples were grown using standard flux methods. The samples grown at the University of Washington were grown by mixing reagents in Canfield crucible sets [https://www.tandfonline.com/doi/full/10.1080/14786435.2015.1122248] then sealing these crucible sets under vacuum in quartz tubes. The Te-flux grown crystals were synthesized using the procedure outlined in Ref.[20] and the Sb-flux grown crystals were synthesized using the procedure outlined in Ref.[54]. The samples grown at Rice University were synthesized from a similar Te flux as described in Ref.[18]

MnTe(0001) epilayers of 50 nm thickness were grown by molecular-beam epitaxy (MBE) on single-crystalline terminated InP(111) substrates using elemental Mn and Te sources. The less than 0.4% lattice mismatch results in single-crystalline hexagonal MnTe growth with the *c*-axis (z-direction) perpendicular to the surface. Two-dimensional growth of α-MnTe is achieved at substrate temperatures of 370–450 °C.

### Transport measurements

Transport measurements were performed on a bulk MnTe sample that was polished and cut with a wire saw to be a bar with dimensions roughly 1 mm × 0.5 mm × 0.1 mm. Gold pads were sputtered onto the sample to make a 4-point measurement. Silver paste and gold wires were used to contact the samples. These measurements were performed in a Quantum Design Dynacool Physical Property Measurement System with standard lock-in techniques.

### Elastocaloric Measurements

Uniaxial stress measurements were performed by gluing a bulk MnTe sample across the gap of a home-made three piezoelectric strain cell which is modeled after the commercial Razorbill CS-100 strain cell. The sample was secured to the titanium cell across a gap of 0.5 mm using Stycast 2850FT Epoxy. An AC voltage of 5 V root mean squared was applied to the inner piezoelectric stack of the strain cell which corresponded to creating an AC displacement of the sample of approximately 0.01% of its length. The temperature fluctuations of the sample induced by this AC strain were measured using a home-made Type E thermocouple comprised of chromel and constantan wires that were silver pasted to each other and the sample. The AC frequency was experimentally determined by measuring the elastocaloric signal at 300K for frequencies in the range of 1−100 Hz, and choosing the frequency with the largest response. This frequency was at the peak of the relevant thermal transfer function, which did not observably shift in the temperature range measured.

**Atomic-scale STEM imaging**

TEM lamellae along different orientations were prepared from two types of bulk MnTe crystals and thin film crystals using a ThermoFisher Scios 2 DualBeam FIB system, with a final polishing process at 2 kV. To minimize the milling damage on the top surface, 2 μm thick platinum or carbon protective layer was deposited on the top of the lamellae. To minimize the surface oxidation and ion implantation, all TEM lamellae were further thinned down to an electron-transparent thickness utilizing a Fischione's Model 1040 NanoMill TEM specimen preparation system at an operating voltage of 900 V and a beam current of 140 pA.

Atomic resolution *Z*-contrast HAADF-STEM imaging was carried out on a Nion UltraSTEM™ 100 microscope, equipped with a spherical aberration corrector, operating at 100 kV and 60 kV, respectively. The HAADF images were acquired with a convergence semi-angle of 32 mrad and collection semi-angles from 80 to 200 mrad. To minimize sample drift effects, we recorded each STEM dataset as 20 successive image frames using a fast scan (1 μs pixel$^{-1}$ dwell time), followed by post-processing drift correction through image registration. Structural distortions were quantified from the STEM-HAADF images by measuring the atomic distances and atomic offsets from their geometric centers. Atomic positions were determined by fitting 2D Gaussian peaks to Mn and Te columns with subpixel accuracy.

STEM simulations to compare with experimental data were performed using the multi-slice method as implemented in *ab*TEM[55]. To match the experimental conditions, we performed the simulations using an aberration-free probe with an accelerating voltage of 100 kV and a convergence semiangle of 32 mrad. Thermal scattering was included through the frozen phonon model with 20 frozen snapshots of the atomic model. The sample thickness was set to 16 nm, and the defocus value was set to 10 Å to obtain good agreement with the experimental data.

**Electron energy loss spectroscopy measurements**

The core-loss electron energy loss spectra (EELS) were acquired using a Nion Iris spectrometer and Dectris ELA high-speed electron-counting detector attached to the Nion UltraSTEM™ at Oak Ridge National Laboratory (ORNL). Spectrum images were acquired using a pixel dwell time of 10 ms, an energy dispersion of 0.9 eV per channel, and a collection semi-angle of 40 mrad. The scanned area was 6 × 6 $nm^2$ with a pixel size of 0.027 nm. The spectrum images were denoised using PCA by selecting the first 20 components according to the scree plots. Te $M_{4,5}$ and Mn $L_{2,3}$ edges were used for elemental mapping. A power-law fitting was used to model the background prior to the core-loss signal. The spectrum images were processed using the Digital Micrograph software.

### EMCD measurements

To minimize the possible electron beam damage, we performed the atomically resolved EMCD experiments at 60 kV with a convergence semi-angle of ~30 mrad and a beam current of 15−20 pA. The on-axis EELS aperture with a collection semi-angle of 25 mrad was used for spectra acquisition. Atomically resolved EELS spectra were acquired at Mn $L_{2,3}$ edge over a 5 nm × 5 nm field of view with 180 × 180 pixels, using a pixel dwell time of 2.5 ms and an energy dispersion of 0.1 eV per channel. 10 successive spectrum images were aligned and integrated to improve the signal-to-noise ratio. Post-processing details of EMCD signal extraction, including background subtraction and post-edge normalization, are described in **Section 8** of *Supplementary Materials*.

### EMCD simulation

We calculated inelastic scattering intensities by using mats.v2 software, which combines the multislice and Bloch-waves approaches for calculating the double differential scattering cross section[56]. The signal distributions of Mn $L_{2,3}$ edges in the real space were calculated with magnetic spins oriented along the different crystallographic directions (see details in **Section 7** of *Supplementary Materials*). To match the experimental conditions, the acceleration voltage was set to 60 kV and the convergence semi-angle of the incident beam to 32 mrad. The sample thickness was set to 18 nm. The inelastic signals were sampled across scattering angles ranging from −25 mrad to 25 mrad in both theta_*x* and theta_*y* directions, with a step size of 2 mrad. The convergence parameter was set to $5 \times 10^{-6}$.

### In-situ heating STEM experiment

STEM cross-sectional FIB samples for in-situ heating experiments were performed using a Thermo Fischer Scientific Helios 5 CX FIB/scanning electron microscope (SEM) DualBeam system at University of Illinois Chicago. STEM lamellae were prepared and subsequently transferred to Fusion Select heating E-chips (E-FHDC-VO-10, Protochips Inc.). In-situ heating STEM experiments were carried out at University of Illinois Chicago using an aberration-corrected JEOL JEM-ARM200CF microscope. The microscope is equipped with a cold-field emission gun and a CEOS aberration corrector and was operated at 200 kV. A probe convergence semiangle of 30 mrad was used to perform atomic-resolution HAADF imaging, with collection angles of 90–370 mrad. A Protochip Fusion Select heating holder with double-tilting capability was used for the in-situ heating experiments. The MnTe specimen was heated to 100 °C at a controlled rate of 30 °C/min. STEM imaging started after reaching the targeted temperature for 30 minutes. To enhance signal quality, atomic-resolution images were recorded as 10 successive frames, subsequently aligned and integrated.

### PFM measurements

We performed piezoresponse force microscopy (PFM) measurement on a bulk MnTe crystal using an Oxford Instruments Cypher-ES atomic force microscope (AFM). We used dual amplitude resonance tracking (DART) to actively monitor the shift in resonance frequency[57,58]. The images are acquired in contact mode using a Pt-coated cantilever (Mikromasch HQ:NSC15, f~325 kHz, k~40 N/m), with a typical AC voltage amplitude of 5 V driven at the second resonance frequency. In DART, the contact resonance frequency is actively tracked during scanning to account for shifts in the contact resonance from point to point; here, the contact resonance is typically in the range of ~900 to 1400 kHz; during scanning, the tracked contact resonance shifts by ~5 kHz. Point spectroscopy data were acquired using switching spectroscopy mode, where a square wave is applied on top of a sawtooth wave to modulate the response of the material[59]; in this case, no DC voltage offset was applied. For each spectrum, at least 25 cycles of 5 V amplitude data were taken and plotted together (not averaged). All spectra were processed in Igor Pro, and images processed in a combination of Igor Pro and Python.

### DFT calculations

All density functional theory (DFT) calculations were performed using the Vienna Ab initio Simulation Package (VASP)[60]. The projector augmented-wave (PAW) method[61] and the Perdew–Burke–Ernzerhof (PBE) generalized gradient approximation[62] were employed. A plane-wave energy cutoff of 500 eV and a $\Gamma$-centered 8 × 8 × 5 $k$-point mesh were used were used for self-consistent-field calculations. For the $P6_3/mmc$ crystal structure, the experimental lattice parameters of $a$=$b$=4.134 Å and $c$=6.652 Å were used[63]. For the $Amm2$, $Cmc2_1$ and $Cm$ crystal structures, the experimentally extracted lattice parameters and atomic coordinates were adopted without further relaxations. For Mn-3$d$ orbitals, onsite Hubbard corrections were applied using $U$=4 eV and $J$=0.97 eV[63]. For all four structures, the calculated atomic magnetic moments are aligned along the $y$-axis. In the band structure calculations with SOC included, the $\langle S_z \rangle$ component is projected onto the band dispersion along the high-symmetry path -$K$–$\Gamma$–$K$, while the $\langle S_y \rangle$ component is projected onto the band dispersion along -$\overline{M}$–$\overline{\Gamma}$–$\overline{M}$.

The distortion modes of α-MnTe structure were investigated using the Bilbao Crystallographic Server[64]. The MSGs and SSGs of the distorted structures were determined using the FINDSPINGROUP[65].

**Acknowledgements**

The electron microscopy work was supported by the U.S. Department of Energy (DOE), Office of Science (SC), Basic Energy Sciences, Material Sciences and Engineering Division, Electron and Scanning Probe Microcopies Program, FWP 83244. Additional support received from NSF through the University of Washington Molecular Engineering Materials Center, a Materials Research Science and Engineering Center (DMR-2308979). The Microscopy work was conducted as part of a user project at the Center for Nanophase Materials Sciences (CNMS), which is a DOE Office of Science User Facility using instrumentation within ORNL's Materials Characterization Core provided by UT-Battelle, LLC, under Contract No. DE-AC05-00OR22725 with the DOE and sponsored by the Laboratory Directed Research and Development Program of Oak Ridge National Laboratory, managed by UT-Battelle, LLC, for the U.S. Department of Energy. A.S.T and R.F.K. were supported by the Office of Basic Energy Sciences, U.S. Department of Energy, award number DE-SC0025396. This work made use of the ThermoFisher Helios 5CX (cryo) FIB-SEM instrument in the Electron Microscopy Core of UIC's Research Resources Center, which received support from UIC, Northwestern University and ARO (W911NF2110052). The atomic force microscopy imaging work performed by R.G. and D.S.G. was supported by the U.S. Department of Energy, Office of Basic Energy Sciences, Division of Materials Sciences and Engineering under Award DE-SC0013957. Part of this work was conducted at the Molecular Analysis Facility, a National Nanotechnology Coordinated Infrastructure (NNCI) site at the University of Washington, which is supported in part by funds from the National Science Foundation (awards NNCI-2025489, NNCI-1542101), the Molecular Engineering & Sciences Institute, and the Clean Energy Institute. P.M.Z. acknowledges funding by the Swedish Research Council under grant number 2024-06617. T.J. and F.K. acknowledge the support by the Ministry of Education of the Czech Republic Grant No.CZ.02.01.01/00/22008/0004594, ERC Advanced Grant No.101095925. and Lumina Quaeruntur fellowship LQ100102602. The CzechNanoLab project LM2023051, funded by MEYS CR, is gratefully acknowledged for the financial support. J.R. and J.Á.C.R. acknowledge the support of the Swedish Research Council (grant no.2025-04514) and Knut and Alice Wallenbergs' foundation (grant no.2022.0079). The simulations were enabled by resources provided by the National Academic Infrastructure for Supercomputing in Sweden (NAISS), partially funded by the Swedish Research Council through grant agreement no.2022-06725, which is also acknowledged for awarding this project access to the LUMI supercomputer, owned by the EuroHPC Joint Undertaking and hosted by CSC (Finland) and the LUMI consortium, where a part of the simulations have been performed. P.D. is supported by US DOE BES DE-SC0026179.

**Data and materials availability:**

Details for the synthesis and characterization of all materials as well as theoretical calculations are described in *Methods* section. The data and scripts that support the findings of this study are available at online database Zenodo[].

## Contribution

G.R. and J.C.I. conceived the project and designed the experiments; J.M.D. and J.H.C. synthesized the single crystals of MnTe using Te-flux and Sb-flux growth at University of Washington; S.X., Z.L. and P.D. synthesized the single crystals of MnTe using Te-flux growth at Rice University; R.C. and P.W. carried out MBE growth of MnTe thin film at University of Nottingham; J.M.D., Y. X. and J.H.C. performed temperature-dependent transport measurements; G.R. carried out the STEM-HAADF and STEM-EELS measurements and their analyses supervised by J.C.I.; X.W.Z. and D.X. carried out the first-principles DFT calculations; R.G. and D.S.G. performed the PFM measurements; J.Á.C.-R. and P.Z. performed STEM imaging simulation and J.R. performed STEM-EMCD simulation; N.K. prepared the STEM sample on heating chips and A.S.T. carried out the in-situ heating STEM measurements under the supervision of R.F.K.; T.J., F.K. and J.M. contributed to the results discussion and performed additional STEM imaging; G.R. and J.C.I. drafted the manuscript with edits from all authors.

## Ethics declarations

The authors declare no competing interests.